\documentclass[aps,pra,twocolumn,floatfix]{revtex4-1}

\usepackage{amssymb,amsmath,amsfonts}

\usepackage{graphicx}
\usepackage{epsfig}

\usepackage[FIGBOTCAP]{subfigure}

\begin{document}

\title{ Interatomic Coulombic Decay in two coupled Quantum Wells   }

\author{{Tamar Goldzak$^{(1)}$,  Liron Gantz $^{(4)}$, Ido Gilary$^{(2)}$,Gad Bahir $^{(4)}$ and Nimrod Moiseyev $^{(1,2,3)}$} \break
$^{(1)}$RBNI- Russell Berrie Nanotechnology Institute, $^{(2)}$Schulich Faculty of Chemistry, $^{(3)}$Physics Center for Nonlinear Physics of Complex Systems,$^{(4)}$ Department of Electrical Engineering, at the  Technion-Israel Institute of Technology, Haifa, 32000, Israel}

\begin{abstract}

Interatomic coulombic decay (ICD) is a relaxation process induced by electronic correlation. In this work we study the ICD process in a two coupled Quantum wells (QWs) nano-structure. We study a simple one-dimensional effective potential using experimental parameters of the semiconductor QW layers i.e. using the single band effective-mass approximation . In our calculations we consider the discontinuity of the effective mass of the electron in each of the QW layers. We control the ICD lifetime by changing the distance between the two wells. The expected overall trend is a decrease of ICD lifetime with a decrease in the distance between the wells. We show that the distance can be tuned such that the emitted ICD electron is trapped in a meta-stable state in the continuum i.e. a one electron resonance state. This causes the life time of the ICD to be an order of magnitude smaller even in very long distances, and improves the efficiency of the ICD. For the ICD to be dominant decay mechanism it must prevail over all other possible competitive decay processes. We have found that the lifetime of the ICD is on the timescale of picoseconds. Therefore, based on our results we can design an experiment that will observe the ICD phenomenon in QWs nano-structure for the first time. This work can lead to designing a wavelength sensitive detector which is efficient even in low intensities.

% In our results we found we can design our system (by changing the distance between the wells) such that the ICD lifetime is an order of magnitude smaller then the expected value.

%The distance between the two wells changes the lifetime of the ICD process. As the distance between the wells decreases the ICD lifetime should also decrease. We design a structure such that the ICD process is enhanced even at long distances between the two wells. The ionized ICD electron is temporarily trapped in a one electron shape type resonance. This resonance state introduces an enhancement in the density of the continuum states at the energy of the ionized electron, resulting a large drop in the ICD lifetime. Based on our results we can design an experiment that will observe the ICD phenomenon in semiconductors nano-structure for the first time. This work can lead to designing a wavelength sensitive detector which is efficient even in low intensities.

\end{abstract}

\maketitle
\section{Introduction}
Interatomic/intermolecular coulombic decay (ICD) is a very efficient and fast electron relaxation process relying on the correlation between electrons. Such process occurs by passing the excess excitation energy of one electron to another electron in a neighbouring atom or molecule, resulting in the ionization of this electron. ICD was first proposed by Cederbaum and co-workers in hydrogen bonded molecular clusters\cite{ICD_clusters}. Past studies focused on weakly bounded systems such as van der Waals clusters and weakly bounded dimers\cite{ICD_clusters_Ne2}-\cite{ICD_clusters_HF}, like the helium dimer which is the most weakly bound system in nature\cite{He_dimer}.

ICD was observed experimentally on nobel gas clusters and dimers like Ne, Ar and He, and in hydrogen-bonded systems like water molecule dimers \cite{ICD_exp_ne1}-\cite{ICD_exp_sum}. All past studies show that the energy transfer in ICD through electron correlation happens also at extremely long distances. This can occur due to the fact that the ionized electron has a long De-Broglie wavelength so its wavefunction couples to the bound states involved in the process.

There are few ways to trigger the ICD process. It could be produced directly from photo-ionization of an inner-valence electron \cite{ICD_clusters_Ne2} or as a result of multistage process such as photo-ionization followed by Auger ionization \cite{ICD_exp_Ar} \cite{Auger_ICD}. Recently another multistage resonant-Auger-driven ICD was proposed which does not involve photo-ionization of inner-valence electron but requires just the excitation of this electron to an unoccupied orbital \cite{res_Auger_ICD}. This special process can yield a very high sensitivity to the location and energies of the ICD electrons, and can be studied in big molecules such as proteins and DNA. It was shown both theoretically and experimentally that the inter-atomic decay rate is strongly dependent distance of neighboring atoms\cite{ICD_clusters_Ne1,ICD_clusters_Ne2,ICD_exp_ne3}. In this sense, the ICD is more efficient as the distance between the atoms is smaller. Moreover it was shown that the ICD lifetime decreases as the number of neighbors increases \cite{ICD_num_neighbor1} \cite{ICD_num_neighbor2}.

Recently it was shown that coupled quantum dots can undergo ICD. Quantum dots (QDs) are solid structures composed of semiconductors which confines electrons in three dimensions and as such they serve as an artificial atoms \cite{QD_atom1} \cite{QD_atom2}. The ICD process in QDs was proven to be very efficient, in comparison to other decay mechanisms exists in the QD, having lifetime of picosecond and less  \cite{ICD_QD1}-\cite{ICD_QD3}. It was shown that the ICD lifetime in QDs grows with the distance between the dots. Quantum wells (QWs) and QDs are widely used in optoelectronic devices such as laser diodes and photo-detectors. Compared to QDs, QWs are easier to grow and control its dimensions.

In this work, we study the ICD decay process in a nano-structure composed of two coupled QWs with different widths, the ICD works in the same manner like in molecular clusters and QDs. An excited electron in one well passes its extra energy to the electron in the neighboring well which is ionized. Excited electrons in QWs have many relaxation pathways, such as spontaneous photon emission, and interaction with phonons. These processes are competing with the ICD in our system. For the ICD to be the dominant decay process it must have lifetime on the same timescale of the shortest decay process in the system. Inter-band spontaneous photon emission is not the most efficient decay process in QW's, and has lifetime on the time scale of nsec \cite{QW_book1}. Scattering of electron with longitudinal optic (LO) phonons i.e. inter-subband relaxation through vibrational modes of the lattice is the most efficient competing relaxation process with lifetime in the timescale of picosecond \cite{QW_book1}-\cite{phonon_QW3}.

 In our calculation we use an effective one-dimensional potential which takes into account only one conduction band of the QWs nano-structure i.e. we use the single band effective mass approximation. We compared the bound state energies and wave functions of this effective one electron hamiltonian with the bound states calculated from the $k\cdot p$ method  \cite{K_P1}-\cite{K_P5}. The $k\cdot p$ method takes into account contribution of eight bands from the conduction and valence bands. Each QW produces a rectangular potential well in the conduction band, which its depth is calculated from experimental measurements of the band gaps of each layer in the nano-structure \cite{QW_parameters1,QW_parameters2}.

We also take into account the change in effective mass of the electron. This effective mass is not constant over the entire space but depends on the semiconductor layers that produces the QWs nano-structure. Here, for the first time, the real band structure and the discontinuity of the effective mass are taken into account in the ICD calculation. Furthermore, our QW nano-structure can be grown in the lab in an easy manner compared to QD systems. In this context the QWs widths and the distance between the wells can be controlled during the growth process. Due to the fact that the parameters of the QW nano-structure in the calculation are taken from real semiconductor properties, one can carry out an experiment based on the results of this paper. A brief description of such experiment will be given below.

We study the ICD process here as a two-electron resonance function which has a finite lifetime. We calculate the life-time of the ICD resonance states at different distances between the neighboring wells. We report an unexpected result, showing that although the wells are far apart from each other the lifetime of the ICD resonance is an order of magnitude shorter compared with the expected value. This unexpected short lifetime enhances the efficiency of the ICD process over other competitive relaxation processes even when the wells are far apart and tunneling is unlikely to occur in these distances. We explain the result using the one electron resonance energies of the ionized electron \cite{NH_QM}. This one-electron resonance originates from the rectangular shape of the two QWs, and introduces a large density of continuum states which makes the two-electron ICD process efficient and fast.

The paper is organized as follows: In Sec.\ref{sec:schem} we introduce a schematic representation of the ICD in the double QW nano-structure. We then move to the methods of calculating the ICD lifetime and a proper representation of the hamiltonian in Sec.\ref{sec:methods}. In Sec.\ref{sec:result} we present our finding and explain our results before concluding in Sec.\ref{sec:conc}.

\section{Schematic representation of the ICD process in the double quantum well nano-structure}
\label{sec:schem}
\begin{figure}[h!]
\begin{center}
\includegraphics[width=1\columnwidth, angle=0,scale=1,
draft=false,clip=true,keepaspectratio=true]{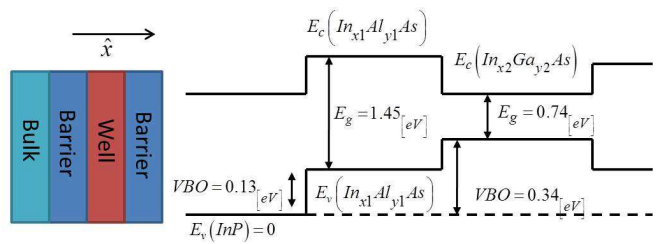}
% \includegraphics[width=\textwidth, angle=0,scale=1,
%draft=false,clip=true,keepaspectratio=true]{QW_new.eps}
\caption{QW nano-structure composed of two different semiconductor materials thin layers. we have chosen the growth direction to be in along the $x$ axis. The semiconductor with the smaller band gap is sandwiched between two semiconductors with a larger band gap. This spacial structure transformed a well in the conduction band, and a barrier in the valence band of the nano-structure. The semiconductor with the smaller band gap is In$_{x2=0.53}$Ga$_{y2=0.47}$As and has a band gap of $ E_{g}=0.74$eV. The semiconductor with the larger band gap is In$_{x1=0.52}$Al$_{y1=0.48}$As and has a band gap of $ E_{g}=1.45$eV. The effective mass of the electron in In$_{0.53}$Ga$_{0.47}$As and In$_{0.52}$Al$_{0.48}$As is $m_{ef}^w=0.045m_e, m_{ef}^b=0.075m_e$ respectively \cite{QW_parameters1,QW_parameters2}.}
\label{Fig::Qwell}
   \end{center}
\end{figure}

The ICD process we study in this work is based on two electrons in two coupled QWs. A schematic presentation of one QW nano-structure is shown in Fig.\ref{Fig::Qwell}, this QW consists of semiconductors and can be grown in the lab. By attaching different layers of semiconductor materials such that a semiconductor with a smaller band gap is sandwiched between two semiconductor  with a larger band gap a QW structure is created (see Fig.\ref{Fig::Qwell}). Due to the difference in the band gaps of each material in the different layers a well structure is formed in the conduction band, and a barrier is formed in the valence band.

 In our system the semiconductor with the smaller band gap is In$_{0.53}$Ga$_{0.47}$As. This semiconductor has band gap of $ E_{g}=0.74$eV and the effective mass of the electron in this layer is $m_{ef}^w=0.045m_e$. The material with the larger band gap is In$_{0.52}$Al$_{0.48}$As, which has a band gap of $ E_{g}=1.45$eV and the effective mass of the electron in this layer is $m_{ef}^b=0.075m_e$ \cite{QW_parameters1}. Both materials are lattice matched to $InP$, this eliminates all the strain effects in the system.

\begin{figure}[h]
\centering
\subfigure[Schematic presentation of the valence and conduction bands of a double QW nano-structure composed of the materials shown in Fig.\ref{Fig::Qwell}. To initiate the ICD process we need to excite the two electrons from the valence band to the conduction band. Each electron is excited with a different laser frequency $\omega_1, \omega_2$ ]{
   \includegraphics[width=1\columnwidth, angle=0,scale=1,
draft=false,clip=true,keepaspectratio=true]{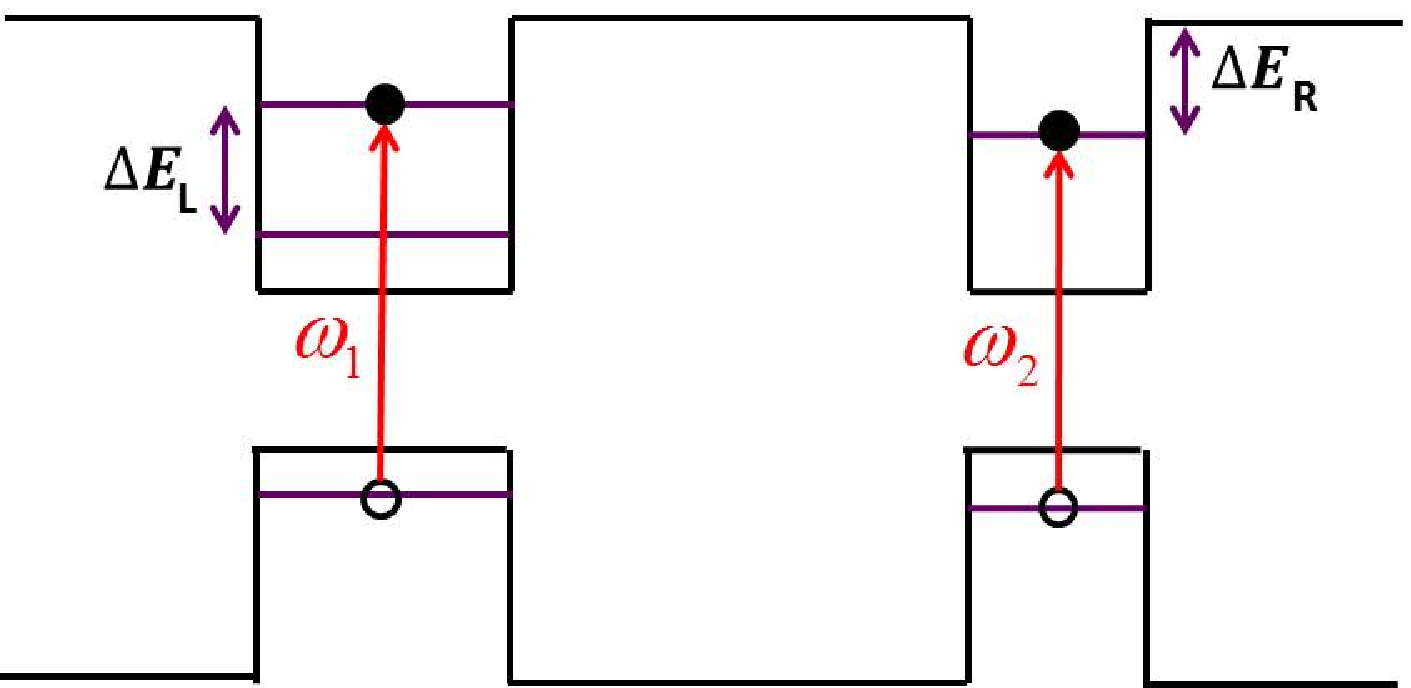}
    \label{Fig:step1}

}
\subfigure[Here we focus on conduction band of the double QW nano-structure. Due to the excitation, one electron is in an excited state located on the left well and the second electron is in the bound state located on the right well.]{
   \includegraphics[width=1\columnwidth, angle=0,scale=1,
draft=false,clip=true,keepaspectratio=true]{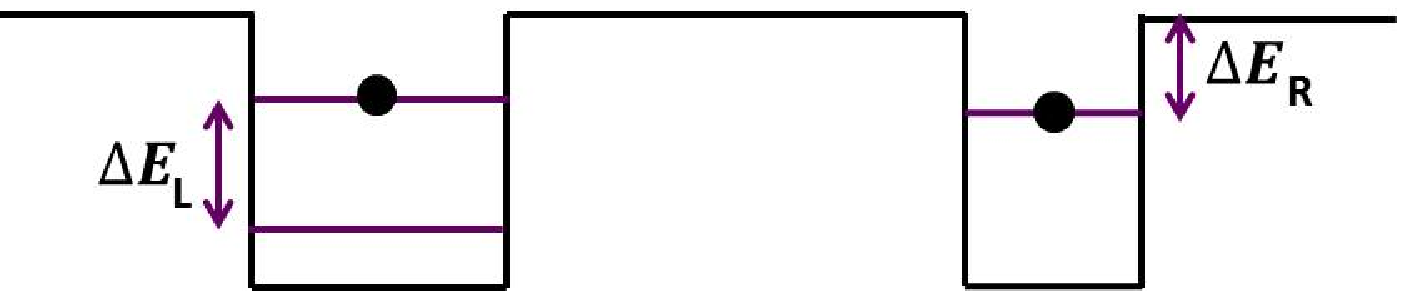}
    \label{Fig:step2}
}
\subfigure[ Due to the correlation between the electrons, the ICD process occurs. The electron in the left well transferring its extra energy to the electron in the right well which is ionized. ]{
    \includegraphics[width=1\columnwidth, angle=0,scale=1,
draft=false,clip=true,keepaspectratio=true]{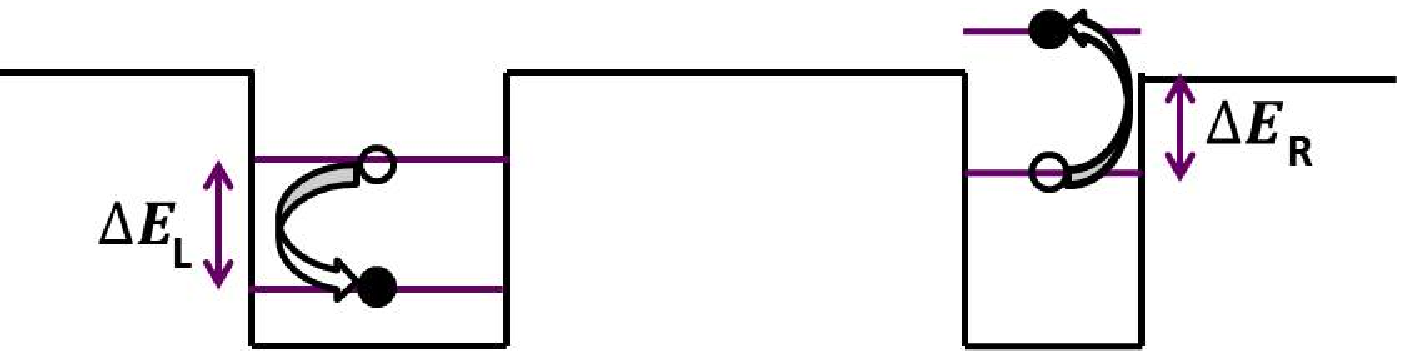}
    \label{Fig:step3}
}
\caption[Optional caption for list of figures]{The ICD process in the two coupled quantum wells nano-structure \subref{Fig:step1}, \subref{Fig:step2} and \subref{Fig:step3}}
\label{Fig:ICDDQW}
\end{figure}

The nano-structure we study of two coupled QWs is shown schematically in Fig.\ref{Fig:ICDDQW}. The left QW is wider than the right QW and supports two bound states , while the right QW supports only one bound state. The ICD process in this system is based on the correlation between two electrons, each one in a different QW. To initiate the ICD process we excite two electrons from the valence band to the conduction band.
We propose an experiment to measure the ICD process using two pulsed lasers at low temperature. In Fig.\ref{Fig:ICDDQW}\subref{Fig:step1} we see a schematic representation of the conduction band and valence band of the coupled QWs nano-structure. The first laser with a frequency of $\omega_1$ matches the interband transition between the first hole bound state and the first excited electronic bound state located in the wider well (see Fig.\ref{Fig:ICDDQW}\subref{Fig:step1}). The second laser with a frequency of $\omega_2$ matches the interband transition between the first hole bound state and the electronic bound state located in the narrow well (see Fig.\ref{Fig:ICDDQW}\subref{Fig:step1}) . By aligning the two lasers to excite the same spot (perpendicular to the growth direction) on the sample, we can treat the problem as one dimensional.
This first step, proposed in Fig.\ref{Fig:ICDDQW}\subref{Fig:step1}, initiates the ICD process.

 The starting point of the ICD, in which the two electrons are in the conduction band is presented in Fig.\ref{Fig:ICDDQW}\subref{Fig:step2}. One can see that the electron located in the left well is in an the excited state while the other electron is in the bound state located in the right well. As a result of the correlation between the two electrons, i.e. repulsive interaction, the electron in the left well decays to the ground state of the system transferring its excess energy to the electron on the right well which is ionized as a result of the electrons correlation (see Fig.\ref{Fig:ICDDQW}\subref{Fig:step3}). By applying low bias to the QWs nano-structure we should sense the change in the dark current due to the ICD process.

There are few requirement for the ICD process. First, the bound state wave-functions of the electrons in the two wells should not overlap each other. If the bound states overlap then tunneling of an electron from one well to the other can take place. This process can overcome the ICD process. Second, the electronic correlation has to be large enough to make the ICD effective. Since the correlation depends on the distance between the electrons we need the wells to be close enough for the correlation to be effective. Therefore, on one hand the wells need to be far apart to prevent from tunneling to prevail, on the other hand the wells should be close enough to allow effective correlation between the electrons. The third requirement is determined by the conservation of energy in the process. Thus we need the relaxation energy of the electron in the left well to its ground state to be larger than the ionization energy of the electron in the right well. This is shown schematically in Fig.\ref{Fig:ICDDQW} i.e. $\Delta E_L>\Delta E_R$.

\section{Methods}
\label{sec:methods}
The one dimensional effective Hamiltonian we use consists of two electrons in the double quantum well nano-structure shown schematically in Fig.\ref{Fig:ICDDQW}. The interaction between the two electrons is a soft coulombic repulsion. We take into consideration that the electrons in the experiment do not change their momentum in $y$ and $z$ direction. In our calculation we include only the dimension parallel to the layers growth direction $x$ (see Fig.\ref{Fig::Qwell}), in this direction a double QW nano-structure is formed. The Hamiltonian of the system is given by :
\begin{eqnarray}
\label{H-2e}
 \hat{H}(x_1,x_2)&=&\hat{H}_{0}(x_1,x_2)+V_{int}(x_1,x_2)\nonumber \\
 &=&\hat{h}(x_1)+\hat{h}(x_2)+V_{int}(x_1,x_2)
\end{eqnarray}
where the electrons' positions are represented by $x_1,x_2$. The Hamiltonian in Eq.\ref{H-2e} contains a sum of two non-interacting one electron hamiltonians $\hat{h}(x_{1}), \hat{h}(x_{2})$ which are coupled by the interaction between the electrons $V_{int}$. The one electron hamiltonian is given by :
\begin{equation}
\label{H-1e}
\hat{h}(x_i)=\hat{P}(x_i)\frac{1}{m_{eff}(x_i)}\hat{P}(x_i)+V(x_i)
\end{equation}
The first term in Eq.\ref{H-1e} is the kinetic energy term where $ \hat{P}(x_i)$ is the momentum operator. The kinetic energy operator takes into account the effective mass of the electron in the different semiconductor layers of the QW depicted in Fig.\ref{Fig::Qwell} i.e. this effective mass presented in Eq.\ref{H-1e} as ${m_{eff}(x_i)}$ and it is discontinuous in $x$. The second term in Eq.\ref{H-1e} is the potential energy in the conduction band of the double QW nano-structure represented schematically in Fig.\ref{Fig:ICDDQW}\subref{Fig:step2}\subref{Fig:step3}. This is an effective potential that takes into account only one conduction band, it will be given explicitly in the next section. The interaction between the electrons $V_{int}(x_1,x_2)$ in Eq.\ref{H-2e} is the soft coulombic repulsion between the electrons and is given by:
\begin{eqnarray}
\label{V-int}
V_{int}(x_1,x_2)&=&\frac{\tilde{\lambda}}{\sqrt{(x_1-x_2)^2+\alpha exp[-\beta(x_1-x_2)^2]}}\\
&&\tilde{\lambda}=\frac{e^2}{4\pi\varepsilon_{r}}\nonumber
\end{eqnarray}
The interaction strength $\tilde{\lambda}$ in Eq.\ref{V-int} is set from the relative permittivity of the layers. For the semiconductors used in our system (see Fig.\ref{Fig::Qwell}) the relative permittivity is $\varepsilon_{r}\cong10\varepsilon_0 $, where $\varepsilon _{0}$ is the vacuum permittivity \cite{Permittivity}. The last term in the square root in Eq.\ref{V-int} is introduced to avoid the singularity of the potential at  $x_1=x_2$ but to keep the coulombic force when the electrons are far apart.
The ICD process is a decay process which has a finite lifetime. We want to calculate the decay rate of this process and insure that the lifetime is at list on the same timescale as the shortest decay process in the system. We are calculating the decay rate of the ICD process using the Fermi golden rule formula \cite{QW_book1}\cite{QW_book2}, such that it is given by:
\begin{equation}
\label{gammaICD}
\Gamma=\frac{2\pi}{\hbar}|\langle\Psi_f(x_1,x_2)|\tilde{H}|\Psi_i(x_1,x_2)\rangle|^2\rho(E_c)
\end{equation}
The lifetime of the ICD process is given by $\tau=\frac{1}{\Gamma}$.
The functions $\Psi_f,\Psi_i$ are the eigen-state wave functions of the unperturbed hamiltonian $\hat{H}_0$ defined in Eq.\ref{H-2e}. $\Psi_i$ represents the initial step of the ICD where the two electrons are in the bound states of the corresponding wells (see Fig.\ref{Fig:ICDDQW}\subref{Fig:step2}). $\Psi_f$ depicts the final step of the ICD where one electron is in the ground state and the second electron is ionized to the continuum (see Fig.\ref{Fig:ICDDQW}\subref{Fig:step3}). These functions are either symmetric (singlet)or anti-symmetric (triplet) with respect to the exchange of the two electrons. These are formed from the eigen-state wave functions of the one-electron hamiltonian proposed in Eq.\ref{H-1e} and are given by:
 \begin{eqnarray}
 \label{psi-int-fin}
 &&\Psi_i(x_1,x_2)=\frac{1}{\sqrt{2}}[\psi_{b2}^L(x_1)\psi_{b1}^R(x_2)\pm\psi_{b2}^L(x_2)\psi_{b1}^R(x_1)]\\
 &&\Psi_f(x_1,x_2)=\frac{1}{\sqrt{2}}[\psi_{b1}^L(x_1)\psi_{c}(x_2)\pm\psi_{b1}^L(x_2)\psi_{c}(x_1)]\nonumber
 \end{eqnarray}
 $\psi_{b1}^L, \psi_{b2}^L$ are the two bound state wave-functions localized in the left well of the double QW shown in Fig.\ref{Fig:ICDDQW} with energies of $E_{b1}^L,E_{b2}^L$ respectively.  $\psi_{b1}^L$ is the ground state wave function of the QW, and $\psi_{b2}^L$ is the excited state wave function. $\psi_{b1}^R$ is the bound state wave-function localized in the right well with energy of $E_{b1}^R$. $\psi_{c}$ is the continuum state wave function of the electron which is ionized from the right well (see Fig.\ref{Fig:ICDDQW}\subref{Fig:step3}). The energy of this continuum state $E_c$ is determined by the conservation of energy in the ICD process, and is given by:
 \begin{equation}
 E_{b2}^L- E_{b1}^L=E_c- E_{b1}^R
 \label{En-cons}
 \end{equation}
 In the calculation of the decay rate $\Gamma$ we assume that only one electron is ionized to the continuum and its momentum in the $y,z$ direction do not change. Accordingly,  we consider a one electron and one dimension density of continuum states $\rho(E)$. It is calculated at the energy of the ionized ICD electron $E_c$ given from the conservation of energy in the process depicted in Eq.\ref{En-cons}.
In the next section we are going to show the results of the calculations to the lifetime of the ICD process using the Fermi golden rule given in Eq.\ref{gammaICD}.

\section{Results and discussion}
\label{sec:result}
The lifetime of the ICD process can be manipulated  by controlling the physical dimensions of the double QW potential shown schematically in Fig.\ref{Fig:ICDDQW}. The dimensions of the different semiconductors layers in the QWs nano-structure were optimized to increase the yield of the ICD process, i.e. to reduce its competition with other decay processes existing in the system. In our calculations we use the parameters of the conduction bands of the QW nano-structure suggested in Sec.\ref{sec:schem}. The only parameter varied in our calculation is the distance between the two coupled QWs.

To evaluate the lifetime of the ICD process at a specific distance between the wells we first calculated the eigenvalues and eigen-states of the one electron hamiltonian proposed in Eq.\ref{H-1e}, in our calculation we used only one conduction band i.e. single band effective mass approximation \cite{QW_book1}. This effective one dimension potential in this hamiltonian is a piecewise potential of the double QW nano-structure (see Fig.\ref{Fig:ICDDQW}). The potential well depth, and the effective mass of the electrons in the different layers is based on the QW structure proposed in Fig.\ref{Fig::Qwell}. We calculated this hamiltonian's eigenvalues by demanding the continuity of the function and flux at the discontinuity points of the potential. The bound states wave functions were required to be square-integrable functions, and by using the transfer matrix method we calculated also the scattering state wave functions of this hamiltonian \cite{QM1,QM_mesoscopic}. To ensure that the effective one dimensional potential is accurate enough, we compared the bound state energies and wave functions with a calculation done using the $ k \cdot p$ method \cite{K_P1}-\cite{K_P5}. In this method we take into account eight bands from the conduction and valence bands, and calculate the electronic bound state energies and wave functions.

\begin{figure}[h!]
\begin{center}
    \includegraphics[width=1\columnwidth, angle=0,scale=1,
draft=false,clip=true,keepaspectratio=true]{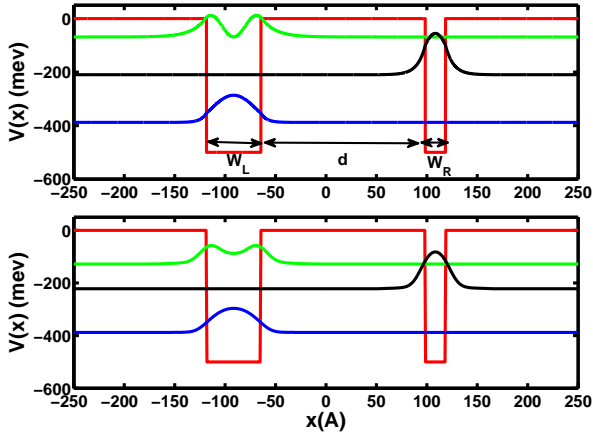}
\caption{(color on line) The one-dimensional effective potential  presenting the conduction band of the coupled QWs nano-structure. The depth of the well is $V_0=-500 $meV, the width of the left and right well respectively is $W_L=53.96$ {\AA}, $W_R=20.05$ {\AA}, the distance between the wells i.e. the barrier width is $d=163 $ {\AA} (solid red-line).The upper panel - shows the calculation of the bound state energies and wave functions using the constant piecewise potential i.e. using the single band effective mass potential. The lower panel - shows the bound states energies and wave functions using the $ k \cdot p$ approximation, which takes into account eight bands from the conduction and valence bands. In both calculations the effective mass discontinuous in $x$. The first bound state wave function located on the left well $\psi_{b1}^L$ (solid blue).The second bound state wave function located on the left well $\psi_{b2}^L$ (solid - green line). The bound state wave function located on the right well $\psi_{b1}^R$(solid-black line) }
\label{Fig::BS}
    \end{center}
\end{figure}
The three bound state wave functions and their energies in the double QW nano-structure are shown in Fig.\ref{Fig::BS} for a specific distance between the wells of $d = 163 {\AA}$. We present in Fig.\ref{Fig::BS}, the bound state wave functions and energies from the the two methods. First is the calculation from the effective one dimension potential (upper panel), this is the bound state wave functions we use to calculate the ICD decay rate. Second, to compare our single band effective mass approximation, we calculated the bound state energies and wave functions using the $ k \cdot p$ method (lower panel). In both calculations presented in Fig.\ref{Fig::BS} the depth of the wells is $V_0=-500 mev$, while the widths of the left and right wells are $W_L=53.96 {\AA}, W_R=20.05 {\AA}$ respectively.

 We obtained in both calculations that the left well supports two bound states while the right well supports only one bound state (see Fig.\ref{Fig::BS}). The ground state wave function $\psi_{b1}^L$ and the bound state wave function located in the right well $\psi_{b1}^R$ in both calculations are almost identical. The excited bound state wave function located in the left well $\psi_{b2}^L$ has a mild difference between the two calculations. The bound state energies are presented in Table.\ref{Table::KP_BS} using the two approaches, $E_b^{cb}$ presents the energies calculated using the one dimension effective potential, and $E_b^{k\cdot p} (mev)$ presents the energies calculated using $ k \cdot p$ method. In the last column of Table.\ref{Table::KP_BS} we present the percentage of the conduction band contribution to the bound state energies and wave functions $\chi_{cb}^{k \cdot p}$  using the $ k \cdot p$ method. By comparing the results from both approaches we obtained that the ground state energies $E_{b1}^L$ are identical. The energies of the bound states located in the right well $E_{b1}^R$ are very close. In these two bound state energies we can see that $\chi_{cb}^{k \cdot p}\cong90$ is very high. There is a bigger difference in the energy of the excited bound state located in the left well $E_{b2}^L$, this is due to the fact that the percentage of the conduction band contribution to that bound state is only $\chi_{cb}^{k \cdot p}=80$.

 \begin{table}
\begin{tabular}{|c||c | c|c |}
  \hline
 $\psi_{bs}$ & $E_b^{cb} (mev)$ & $E_b^{k\cdot p} (mev)$ & $\chi_{cb}^{k \cdot p}$   \\\hline
 $\psi_{b1}^L$  & -388 & -388  & 0.91 \\\hline
 $\psi_{b1}^R$  & -210 & -222 &0.88 \\\hline
 $\psi_{b2}^L$  & -69  & -128 & 0.8 \\\hline
\end{tabular}
\caption{The three bound state energies using the single band effective mass approximation i.e. taking into account only one band from the conduction band of the double QW ($E_b^{cb}$). The bound states energies using the k.p approximation i.e. taking into account eight bands from the conduction and valence bands of the double QW ($E_b^{k.p}$). The percentage of the conduction band contribution to the bound states using the k.p approximation ($\chi_{cb}^{k.p}$) }
\label{Table::KP_BS}
\end{table}

From this comparison we can see that the bound state energies and wave functions are very similar, there is a slight shift in the bound state energy containing more valence band contribution, this changes will not change the result of the ICD decay rate significantly. Furthermore in both methods changing the distance between the wells $d$ did not change the bound state energies. This comparison between the bound states obtained from the $ k \cdot p$ method and the one-dimensional effective piecewise potential has shown that we can use the single band effective mass approximation in the calculation of the ICD decay rate.

\begin{figure}[h]
\begin{center}
    \includegraphics[width=1\columnwidth, angle=0,scale=1,
draft=false,clip=true,keepaspectratio=true]{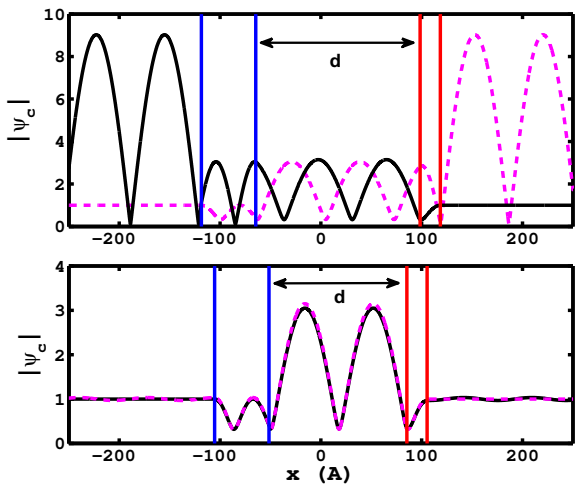}
\caption{(color on line) The continuum state wave functions of the ionized electron $\psi_c$ due to the ICD process. The different panels referring to different distances between the wells in the one electron potential given in Fig.\ref{Fig::BS}. upper panel - the distance between the wells is $d=163{\AA} $, lower panel- the distance between the wells is $d=136.8 {\AA}$. The continuum state has energy of $E_c=109.3 $meV in both panels, this energy is set from the conservation of energy in the ICD process see Eq.\ref{En-cons}. One of the continuum state $\psi_c$ describes the electron arriving from $\infty$ while the other continuum state describes the electron that arrives from $-\infty$ (solid black and dashed magenta respectively). The edges  of the left and right wells in the double QW nano-structure are shown in solid blue and solid red lines respectively.   }
\label{Fig::CS}
    \end{center}
\end{figure}

 In all the calculations of the ICD lifetime all the potential parameters remained constant and only the distance between the wells was varied. We calculated the energy of the ionized electron, i.e. the energy of the continuum state $\psi_c$  in Eq.\ref{psi-int-fin}. The energy of this continuum state is $E_c=109.3 $meV which is derived from the conservation of energy requirement in the ICD process given in Eq.\ref{En-cons}.
Although the bound-state energies and wave functions are hardly changed with the distance between the wells, the spacial structure of the continuum state wave function changes with the distance. In Fig.\ref{Fig::CS} we show the two continuum state wave functions at two different inter-well distances. As one can see the continuum state amplitudes and shapes are changing with the distance between the wells. These changes affect the ICD lifetime, due to the fact that the overlap between the bound state wave functions vanishes and therefore only the continuum state wave functions couples between the bound state wave functions.

\begin{figure}[h]
\centering
\subfigure[The lifetime of the ICD process $\tau$ as a function of the distances between the wells $d$. ]{
   \includegraphics[width=1\columnwidth, angle=0,scale=1,
draft=false,clip=true,keepaspectratio=true]{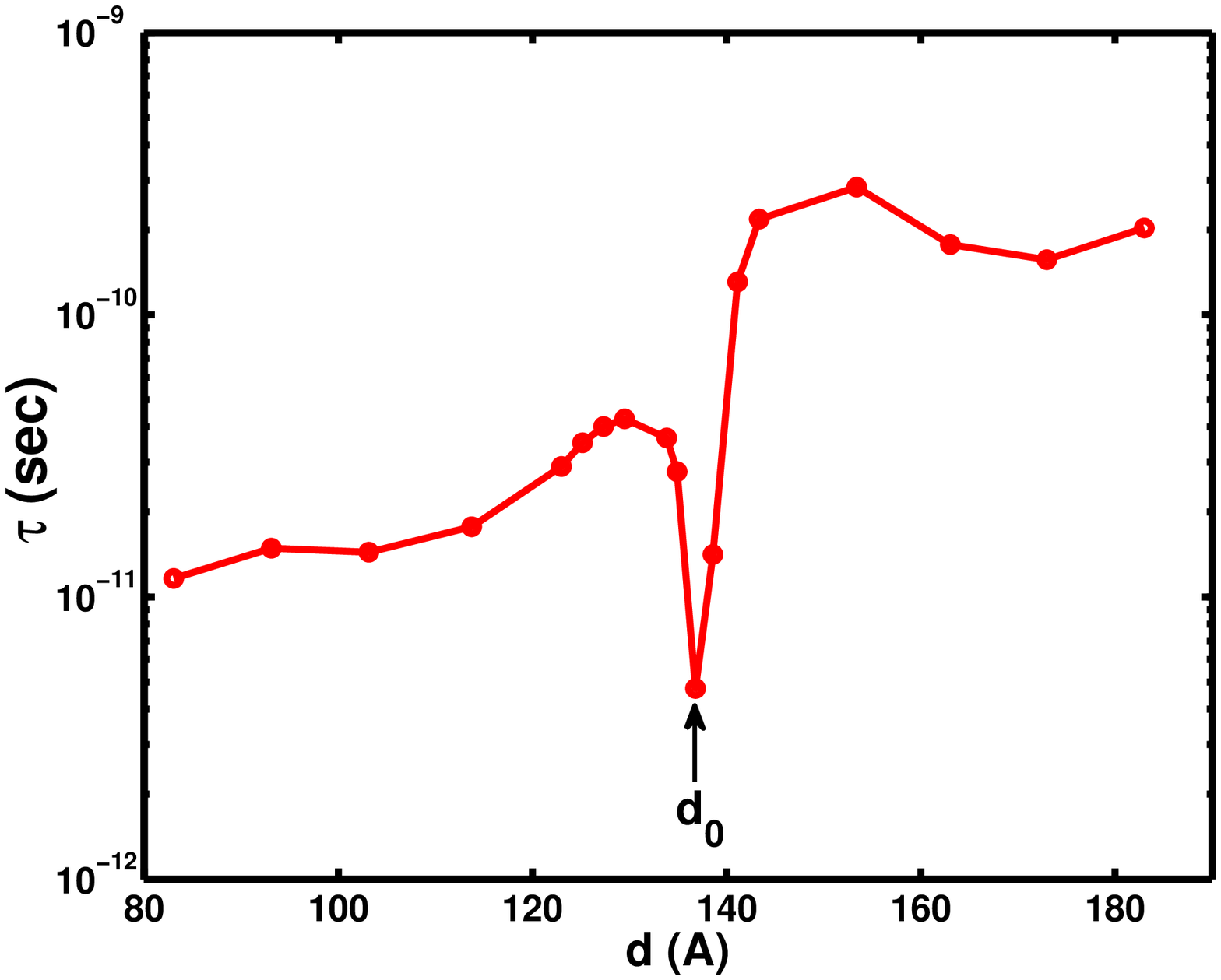}
    \label{Fig:lifetime}

}
\subfigure[The overlaps between bound state wave functions of the two wells (treating each well separately) as a function of the distance between the wells. The overlap of the bound state in the right well and the ground state in the left well (solid red). The overlap of the bound state in the right well and the first excited state in the left well (dashed blue)]{
   \includegraphics[width=1\columnwidth, angle=0,scale=1,
draft=false,clip=true,keepaspectratio=true]{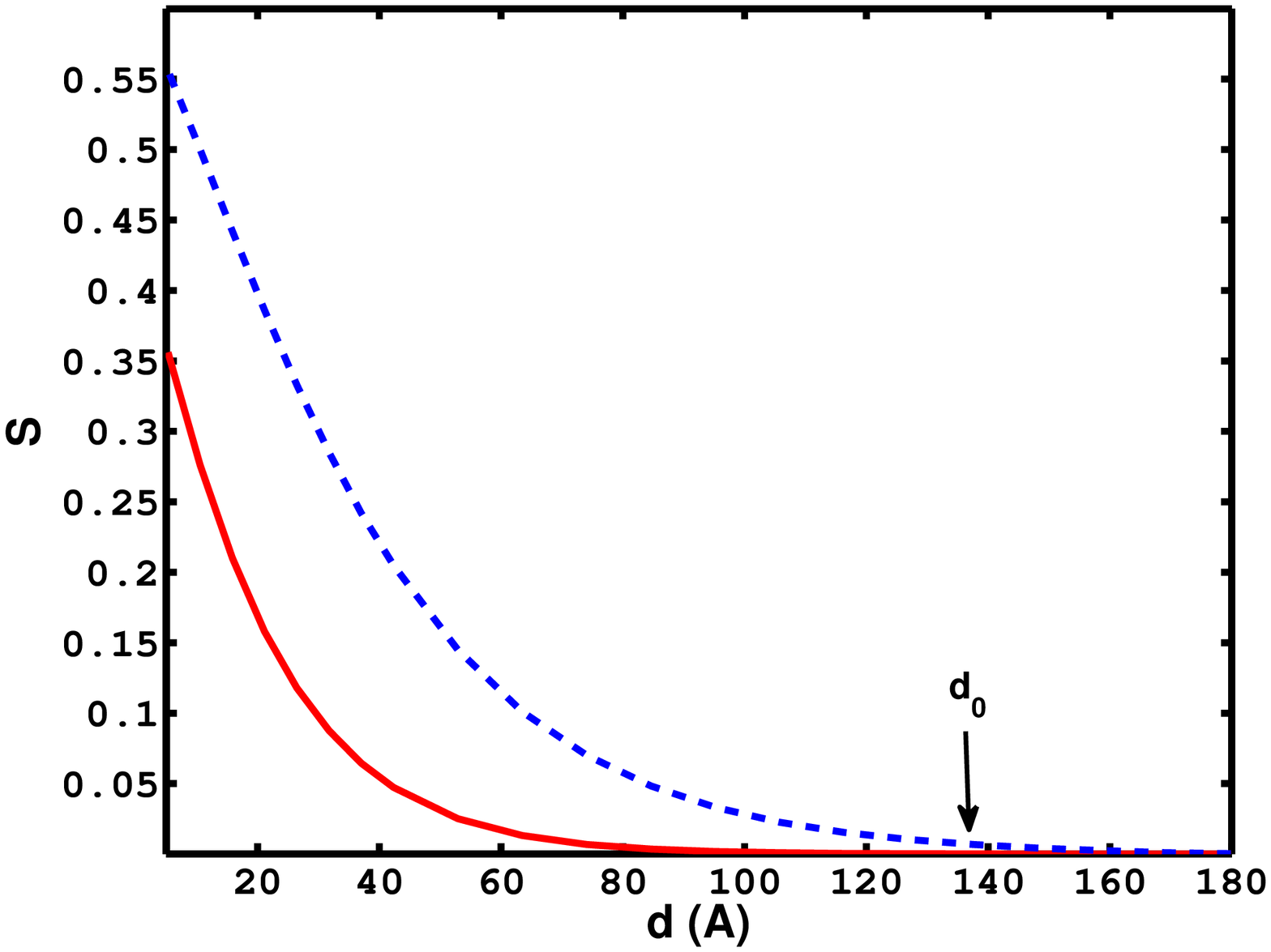}
    \label{Fig:overlap}
}
\caption[Optional caption for list of figures]{The ICD life time is shown in Fig.\subref{Fig:lifetime}, and the overlap between the bound state wave functions in the different wells is shown in Fig.\subref{Fig:overlap}}
\label{Fig:Life}
\end{figure}

We calculated the decay rate of the ICD process using the fermi golden rule formula in Eq.\ref{gammaICD}, while changing the distance between the wells. Here we separated between the singlet and triplet eigen functions of the unperturbed hamiltonian in Eq.\ref{psi-int-fin}. In Fig.\ref{Fig:lifetime} we show the lifetime of the ICD as a function of the distance between the wells using the triplet functions, we got similar results also for the singlet functions. We expect the life time to grow as the distance between the wells increases, due to the fact that as the distance increases the correlation between the electrons decreases and so does the decay rate. In Fig.\ref{Fig:lifetime} we see that the overall trend follows this expectation except around the point of $d=d_0=136.8${\AA}. At this point we see a surprising sharp drop in the lifetime. One can see that although the distance at this point is quite large, we get very short life time of several picosecond. This is an order of magnitude shorter then what is expected.

To show that this is indeed a result of ICD we also calculated the overlap between the bound state wave functions of the right well with those on the left well by treating the wells as separate systems. The results are shown in Fig.\ref{Fig:overlap}. As one might intuitively expect, we see that as the distance between the wells increases the overlap between the bound state wave functions in the two different wells decreases. This overlap is a measure of the tunneling in the system, which means that tunneling is effective at small distances. When tunneling occurs in our system the ICD is not the dominant decay process. Around the special point of $d_0$ the overlap between the bound state wave functions of the two wells is very small so we know that the tunneling is not effective. This is in contrast to smaller distances where the lifetime is short but the overlap is large enough to make tunneling the dominant process.

In order to explain this interesting result, we need to examine the expression for the ICD decay rate given in Eq.\ref{gammaICD}. One possible reason for the increased decay rate could lie in the shape of the continuum wave function in the region between the wells. We can look at the continuum state amplitude of two different distances: at $d_0$ and at $d_1=163${\AA} (see Fig.\ref{Fig::CS}), and observe only a small difference in the amplitudes of these two wave functions. This cannot be the reason for such a significant change in the lifetime.

Another factor that may affect the decay rate is the density of states in the continuum at the energy of the escaping electron $\rho(E_c)$. The density of states in the free one-electron picture follows $\rho_{free}\propto\frac{1}{\sqrt{E_c}}$. If this was the case in our system the trend in the lifetime  should not change because the energy of the escaping electron $E_c=109.3 $meV remains constant with the distance. Due to the structure of the potential, the density of states has peaks which are correlated with this structure. We calculated the density of continuum states for the one-electron hamiltonian in Eq.\ref{H-1e} (used also for the lifetime calculations) at different distances by choosing vanishing boundary condition in a large box $\psi(x=\pm L/2)=0$ \cite{DOS1,DOS2}. We
evaluate numerically the one dimensional density of states in the energy of the continuum states i.e. energies above the threshold, by calculated the following expression in a large box:
\begin{equation}
\label{densityEq}
\rho(E)=\frac{1}{L}(\frac{\Delta E}{\Delta n})^{-1}
\end{equation}
We normalized this density of states by dividing it to the length of the box $L$.
\begin{figure}[h!]
\centering
\subfigure[The density of continuum states for the one electron hamiltonian $\hat{h}(x_i)$  of the system for two different inter-well distances, calculated numerically by solving the problem in a large box $L=15870{\AA}$. The density in a distance of $d=d_0=136.8 {\AA}$, which is the spacial distance where there is a sharp drop in the lifetime of the ICD (see Fig.\ref{Fig:lifetime}) (solid blue). The density for a distance of $d_1=163A$ (dashed red). We can see that that the background of this plot is the density of free particle in one dimension. The peaks in the density are correlated with the structure of the potential, therefore at different distances we see different picture of the density of continuum states and its peaks. The continuum state energy of the ICD electron escaping the system is marked on the plot as $E_{c}$ and it do not change with the distance.]{
   \includegraphics[width=1\columnwidth, angle=0,scale=1,
draft=false,clip=true,keepaspectratio=true]{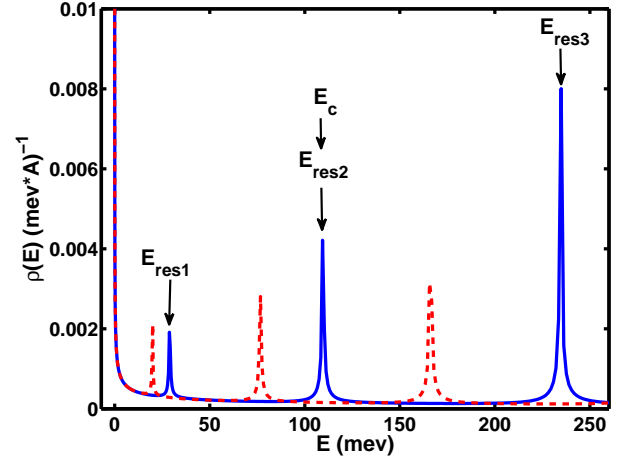}
    \label{Fig:rho}
}
\subfigure[Resonance solutions of the one electron Hamiltonian in Eq.\ref{H-1e} on the complex plane with a distance of $d=d_0=136.8A$ between the wells. One can see the correlation between the resonance position and the peaks in the density of states]{
   \includegraphics[width=1\columnwidth, angle=0,scale=1,
draft=false,clip=true,keepaspectratio=true]{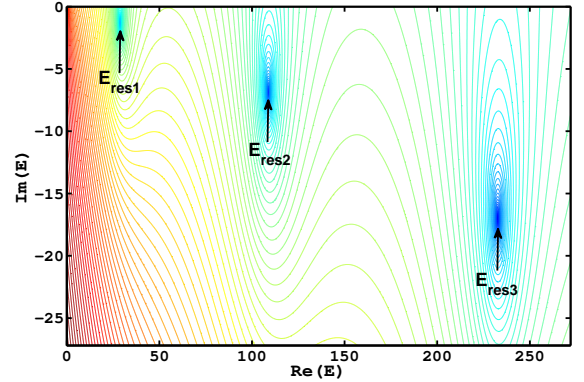}
    \label{Fig:E-res}
}
\caption[Optional caption for list of figures]{\subref{Fig:rho} the density of continuum states for the one electron hamiltonian presented in Eq.\ref{H-1e} in two different inter-well distances. \subref{Fig:E-res} The resonance solutions to the one electron hamiltonian on the complex plain. }
\label{Fig:density}
\end{figure}

Fig.\ref{Fig:rho} shows the density of the continuum states of the one-electron hamiltonian $\hat{h}(x_i)$ as a function of the continuum energy in two different structures, i.e. two distances between the wells. The solid blue line is at the distance $d_0$ , and the red dashed line is at a distance of $d_1=163${\AA}. One can see that although the background remains the same in both plots, the peaks appear at different energies. Furthermore at $d_0$ there is a peak in the density of states at the energy of the ionized ICD electron $E_c$. While in the other distances such as in $d_1$ the energy of the ionized ICD electron which is also $E_c$ there is no peak in the density of states. This means that the decay rate will be an order of magnitude larger than expected in the spacial distance of $d_0$. This explains the sharp drop in the ICD lifetime in Fig.\ref{Fig:lifetime} around the point $d_0$.

One can calculate the resonance states of the one-electron hamiltonian $\hat{h}(x_i)$ given in Eq.\ref{H-1e} by imposing outgoing boundary conditions on the schr\"{o}dinger equation \cite{NH_QM}. The resonance picture of the one electron hamiltonian can help in understanding the density of states picture. The resonance energies on the complex energy plane for inter-well distance of $d_0$ are shown in Fig.\ref{Fig:E-res}. It is evident that the position of the resonance energies matches the peaks in the density of states (see Fig.\ref{Fig:density}\subref{Fig:rho} in solid blue).

This can be explained by the fact that the resonance states are connected directly with the poles of the S-matrix in the complex plane \cite{NH_QM}. When the poles of the S-matrix are isolated from each other, and close enough to the real axis, one can associate the peaks in the cross section with real part of the poles of the S-matrix on the complex energy plane \cite{cross_sec}. The cross section  depends on the density of states via the S-matrix or Green operator \cite{Green}, such that the peaks in the density of states should appear at the energies of the poles, i.e. the positions of the resonance states.

 From this discussion we conclude that the efficiency of the ICD process is enhanced at the distance of $d_0$. This makes the ICD life-time on the same time scale of the dominenet competing decay process in the QWs nano-structure. The reason for this significant enhancement is that the ionized ICD electron is temporarily trapped in a shape type resonance state.  This trapping enables us to get an efficient ICD ionization that competes with other decay processes even at very long distances. Therefore the calculations of the shape type resonances as function of the distance between the two QWs provides us a powerful computational tool for designing an experiment to observe the ICD phenomenon.
\section{concluding remarks}
\label{sec:conc}
In this work we propose for the first time an ICD process in a realistic structure of two coupled QWs. The system we studied is based on real  physical parameters of QWs in semiconductors materials. The one electron effective potential is set from the band-structure of the semiconductor materials in the QW nano-structure. We also take into account the discontinuity in the effective mass of the electrons in the different QW layers, and the permittivity of the layers. The ICD lifetime depends very strongly on the distance between the wells. The overall trend is that by increasing the distance between the wells the lifetime is increasing. Our results shows that the shortest ICD lifetime in our system is of several picoseconds. This means that the ICD is on the same time scale of inter-subband relaxation with LO phonons which dominant inter-subband relaxation process in QW. By designing a sample which matches the ICD conditions and with enough double QW periods, this phenomenon should be observable experimentally.

The main result of this paper shows that the parameters of the potential can be manipulated such that the ICD process is enhanced. The ionized electron is temporarily trapped in a shape type resonance state, this resonance state introduces a peak in the density of continuum states resulting in a very short lifetime of the ICD process on the time scale of picosecond,  even at a very long distance between the wells. Based on our result and understanding of the ICD process, we can design an experiment which will show this phenomenon for the first time in nano-structures. This can lead to designing a photo-detector which is very sensitive to wave-length even at very  low intensities .

\begin{acknowledgments}

The authors acknowledge ISF grant 298/11 and the I-Core: the Israeli Excellence Center "Circle of Light" are acknowledged for their support.

\end{acknowledgments}

\end{document}